# Activating Basal Planes in Transition Metal Dichalcogenides for $CO_2$ Reduction to CO through Alloying

Eric Montufar-Morales,[a,*] Daniel Rinder,[b] Pravan Omprakash,[a] Rohan Mishra,[c,a,*] Gwan Yeong Jung[c,d,*]

[a]Institute of Materials Science & Engineering, Washington University in St. Louis, St. Louis, MO, USA

[b]Department of Computer Science & Engineering, Washington University in St. Louis, St. Louis, MO, USA

[c]Department of Mechanical Engineering & Materials Science, Washington University in St. Louis, St. Louis, MO, USA

[d]Department of Chemistry, Incheon National University, Incheon, 22012, Republic of Korea

*Denotes corresponding authors

E.M.M.: e.b.montufar-morales@wustl.edu; R.M: rmishra@wustl.edu; GYJ: gyjung@inu.ac.kr

**Abstract**

Transition metal dichalcogenides (TMDCs) have emerged as highly tunable platforms for electrocatalysis, particularly for the $CO_2$ reduction reaction ($CO_2$RR). While TMDC edge sites exhibit catalytic activity, the basal plane is typically inert, severely limiting the overall active site density. In this work, we show that sulfur vacancies activate the basal plane of 1H-TMDCs, while concurrent solid-solution alloying provides a mechanism to broadly tune intermediate adsorption energies. We evaluate a library of quasi-binary TMDC sulfide alloys comprising V, Nb, Ta, Mo, and W, screening them by stability, defect energetics, and competitive selectivity to identify the most effective catalysts for $CO_2$RR. Electronic-structure analysis reveals that a *d*-band center closer to the Fermi energy weakens intermediate binding by leaving key bonding states unoccupied above the Fermi energy. Our calculations identify (Nb,Ta)$S_2$ as a promising catalyst with low sulfur vacancy formation energies, near-optimal $CO_2$RR intermediate binding, and selectivity against the hydrogen evolution reaction. Overall, this work establishes a rational design framework for TMDC alloy catalysts through the simultaneous use of defect engineering and alloying.

## 1. Introduction

Electrochemical reduction of $CO_2$ to CO offers an attractive pathway for producing a critical syngas precursor for hydrocarbon synthesis via the Fischer-Tropsch process[1–4]. Metallic Au and Ag serve as conventional catalysts for this reaction. Both selectively reduce $CO_2$ to CO without further reactions[5], whereas metallic Cu can generate more valuable multi-carbon products, albeit with lower selectivity[6]. However, the high cost of Au and Ag, and large overpotentials associated with Cu[7] make industrial-scale implementation challenging[6,8], hence necessitating an efficient and cost-effective catalyst. Two-dimensional (2D) transition metal dichalcogenides ($MX_2$, $X$ = (S, Se, Te)), or TMDCs, have gained attention as effective catalysts for a wide range of chemical reactions such as hydrodesulfurization[9] and hydrogen evolution,[10–13] and more recently for $CO_2$ reduction[14,15]. The ability of TMDCs to accommodate various cation species via alloying provides a compositional flexibility[16,17] that is ideal for tuning catalytic performance. For $CO_2$ reduction, catalytic activity has been largely attributed to TMDC edge sites,[5,15,17–19] as the basal plane has traditionally been considered chemically inert[20]. Activating the basal plane would unlock the dominant surface area of 2D TMDCs, greatly increasing the density of catalytically accessible sites and providing a powerful route to enhance the intrinsic activity for the $CO_2$ reduction reaction ($CO_2$RR).

Efforts to activate the TMDC basal plane have yielded several strategies. One approach involves the introduction of sulfur vacancies (S-vacancies), which have been theoretically predicted and experimentally shown to increase activity for $CO_2$RR[21]. The efficacy of alloying in quasi-binary TMDCs was demonstrated by Hemmat et al., whose atomic imaging confirmed that miscible quasi-binary alloys, such as (Nb, Ta)$S_2$, can outperform conventional catalysts for the $CO_2$RR[16]. More recently, Akhound et al.[22] showed that the catalytic landscape can be further modulated through transition metal alloying by leveraging strain effects arising from differing bond lengths to tune a catalyst's intrinsic activity. While these strategies independently mitigate surface inertness, their synergistic impact on $CO_2$RR selectivity against the competing hydrogen evolution reaction (HER) remains unexplored. However, a key unresolved question is how alloying redistributes adsorption energies around individual vacancy sites, and how those local variations govern competition between $CO_2$RR and HER.

In this Article, we predict (Nb, Ta)$S_2$ to be an active and efficient catalyst for the $CO_2$RR. We arrived at that conclusion through a systematic screening of binary TMDCs, $M_{0.5}N_{0.5}$ $S_2$, where $M, N \epsilon \{\mathrm{Mo, Nb, Ta, V, W}\}$. We find that sulfur vacancies serve as discrete active site for the $CO_2$RR and lower the energy of the reaction on the basal plane, when compared to a pristine basal plane. By alloying, we expand the distribution of binding energies that can exist on the catalyst[21]. We find that among the catalysts calculated, (Nb, Ta)$S_2$ shows narrow distributions in vacancy formation energies, $CO_2$RR, and HER intermediate energies throughout a variety of different active sites. The distributions for (Nb, Ta)$S_2$ show near-zero energies for the CO2RR intermediates, which is optimal for reducing the over potential for the reaction. Balance between the population of bonding and anti-bonding states in (Nb, Ta)$S_2$ gives the optimal bonding strength with intermediates. Our work shows that the combined strategy of alloying and defect engineering can be used to effectively tune the catalytic behavior of TMDCs for the $CO_2$RR towards CO, enabling them to be efficient, selective and cost-effective catalysts.

## 2. Methods

### 2.1. Density-functional theory calculations

To model solid solution quasi-binary TMDC alloys, we generated special quasirandom structures (SQS)[22] using the Alloy Theoretic Automated Toolkit (ATAT)[23]. We used a 108-atom supercell for the defect-free TMDC alloy. All alloy structures contain an equimolar concentration of the two cations. A vacuum spacing of ≥ 19 Å was used along the *z* direction to ensure adequate separation between periodic layers. All calculations were performed using the projector-augmented wave potentials[24] as implemented in the plane-wave DFT code, Vienna ab initio simulation package (VASP)[25,26]. The Perdew-Burke-Ernzerhof (PBE) functional within the generalized gradient approximation (GGA)[27] was used to describe the exchange-correlation interactions. A plane-wave basis set with an energy cutoff of 500 eV and an electronic convergence criterion of $10^{-6}$ eV was used. *K-point*s grid of $2 \times 2 \times 1$ and $3 \times 3 \times 1$ were used for the structural relaxations and static calculations, respectively. The structures were relaxed until the forces on the atoms were < $10^{-2}$ eV/Å. To evaluate the density of states, we used 5000 grid points along the energy axis (NEDOS tag in VASP) with a Gaussian smearing of 0.1 eV.

**2.2. Free-energy calculations**

The enthalpy of mixing[28], $\Delta H_{mixing}$, is given by

$$\Delta H_{mixing} = E^{alloy} - \sum_{i} n_i E_i^{unary}, \tag{1}$$

where, $E^{alloy}$ is the total energy per formula unit of the alloy, $n_i$ is the molar fraction of the $i^{th}$ component, which in this case is a unary TMDC, and $E_i^{unary}$ is the energy per formula unit of corresponding component. The $\Delta H_{mixing}$ is calculated at 0 K in this work.

The free energy of adsorption, $\Delta G_{ads*}$, is determined as

$$\Delta G_{ads*} = \Delta H^{ads*} + \Delta E_{ZPE} - T\Delta S_{ads*} + E_{cor}. \tag{2}$$

Here, $\Delta H^{ads*}$ is the adsorption energy, $\Delta E_{ZPE}$ is the difference in zero-point energy between the adsorbed molecule and the molecule in vacuum, and $\Delta S_{ads*}$ is the vibrational entropy contribution at a temperature, $T$ = 300 K, and $E_{cor}$ is an additional correction term to account for a DFT-overestimation of the total energy of $CO_2$ and $H_2$, having a correction value of 0.4 and 0.013 eV, respectively[17].

The formation energy of a charge-neutral sulfur vacancy, $\Delta E_{vac}$, is calculated using (3):

$$\Delta E_{vac} = E_{def} - E_{perf} + \mu_s. \tag{3}$$

In equation (3), $E_{def}$ is the total energy of the supercell with a sulfur vacancy, $E_{perf}$ is the total energy of the supercell in its pristine state, and $\mu_s$ is the chemical potential of sulfur. $\Delta E_{vac}$ is calculated in a sulfur-rich environment to allow for easy comparison between compositions. This means that $\mu_s$ is defined by

$$\mu_s = \frac{1}{8} E_{S_8}, \tag{4}$$

where $E_{S_8}$ is the DFT energy of an $S_8$ molecule in vacuum. Modeling a sulfur-poor environment would cause overall $\Delta E_{vac}$ to lower, but would not be comparable for different alloys, as a sulfur-poor environment would require chemical potentials of the transition metals in the alloys. In total, we performed around 15 different vacancy configurations for each alloy TMDC, all of which is present in the $\Delta E_{vac}$ analysis. Further analysis used a smaller amount of vacancy configurations for all catalysts except for (Nb, Ta)$S_2$, as that was seen to be the most promising catalyst.

## 3. Results and Discussion

### 3.1. Sulfur vacancies activate unary TMDC basal planes for $CO_2$ adsorption

We consider equimolar TMDCs with transition metals from groups V and VI, including V, Nb, Ta, Mo, and W, because these elements are stable in the 1H phase[29]. We model quasi-binary TMDCs as random solid solutions where the two TMs are distributed randomly on the cation sublattice, as illustrated in Fig. 1a. We then introduce a basal-plane sulfur vacancy by removing one sulfur atom from the 108-atom supercell, indicated by the dashed red circles. The resulting sulfur vacancy serves as an active site for the adsorption of $CO_2$ and $H_2$ molecules.

The $CO_2$RR toward CO can be described in four steps, beginning with gaseous $CO_2$ prior to adsorption, as highlighted in step 1 of Fig. 1b. $CO_2$ then adsorbs on the pristine TMDC surface as a carboxyl group (COOH*) through a proton and electron transfer, either on a sulfur atom, as shown in box $2_p$, or on the transition metal, as shown in box $2_v$ in Fig. 1c. Through a second proton and electron transfer, the adsorbed species forms CO*, as shown in boxes $3_p$ and $3_v$, before the CO molecule desorbs from the surface.

To demonstrate the effect of S vacancies on basal-plane catalytic activity, we evaluate the $CO_2$RR free-energy pathway of pristine $MoS_2$ at a basal S site and at a basal S-vacancy site. The free-energy diagrams for these two cases are shown in Fig. 1b, with the S-vacancy site in pink and pristine $MoS_2$ in blue. The rate-limiting potential, $U_L$, is defined by the largest, positive free-energy change along the reaction pathway and serves as a key descriptor for catalytic activity[30]. For pristine $MoS_2$, the highest energy barrier occurs during the conversion of $CO_2$(g) to COOH*, corresponding to a $U_L$ of 2.16 eV. Introducing an S vacancy lowers this barrier by 1.47 eV, to 0.69 eV. Although the $CO_2(g) \rightarrow COOH^*$ step still remains rate-limiting, the reduced $U_L$ demonstrates that $V_{\dot{S}}$ improve the reaction energetics of $CO_2$ reduction on the $MoS_2$ basal plane[20].

S-vacancies were also modeled in $NbS_2$, $TaS_2$, $VS_2$, and $WS_2$, and we compared their activity with the (111) surface of Ag, a conventional catalyst for CO2RR[2], as shown in Fig. 1d. The $U_L$ values of all five unary TMDCs were lower than that of Ag. $TaS_2$ stands out among the unary TMDCs with an $U_L$ of 0.08 eV compared to 1.29 eV for Ag; however, it can also be seen that both the intermediate steps have weak, positive binding energies of 0.08 and 0.11 eV, whereas other TMDCs, such as $WS_2$ and $MoS_2$, show strong binding energies to CO of -0.11 eV and -0.41 eV, respectively. Therefore, we hypothesize that alloying $MoS_2$ with other transition metals to form quasi-binary TMDCs can tune the adsorption energies of the reaction intermediates, potentially reducing limiting potential, $U_L$, toward zero.

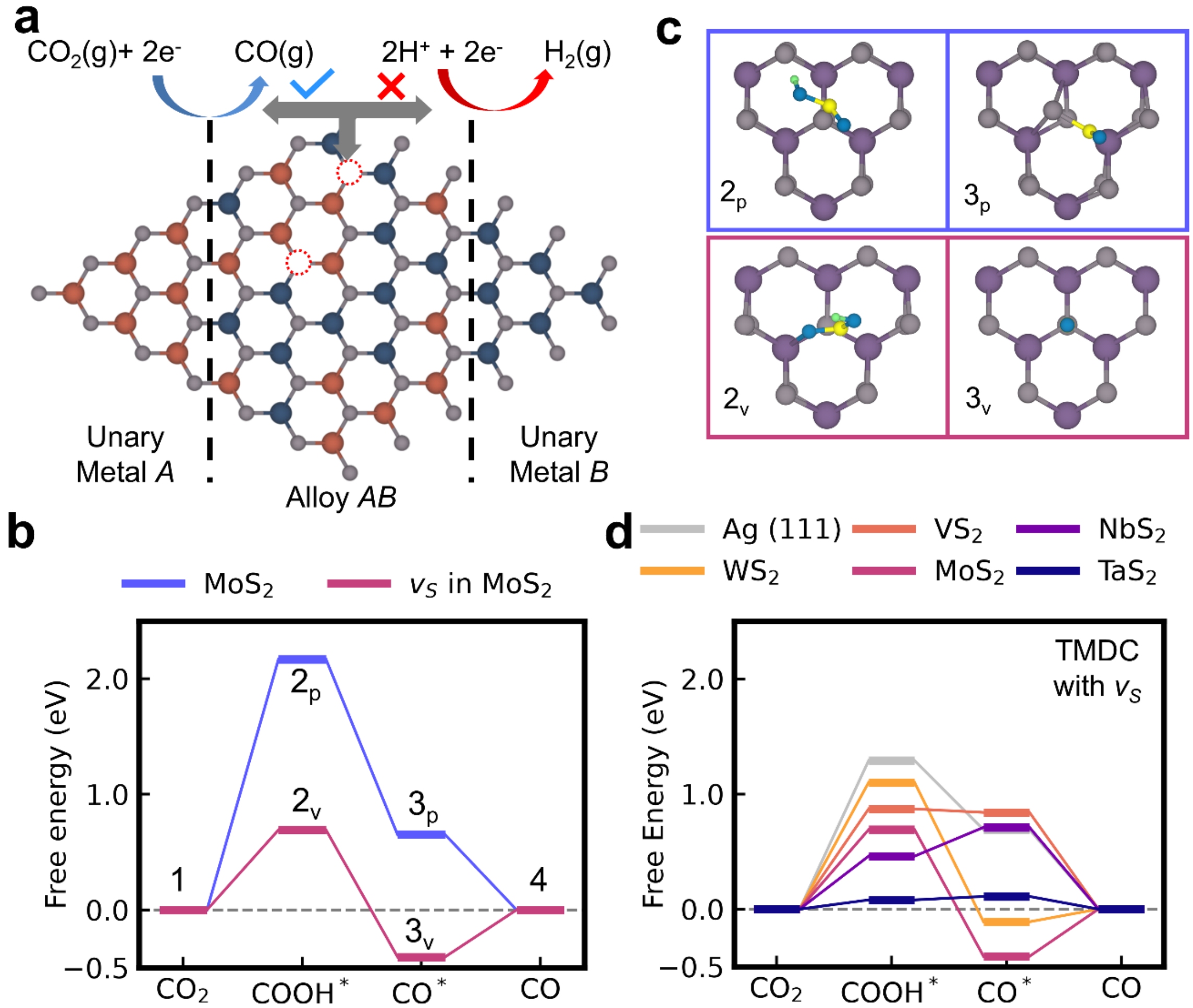

**Figure 1: a)** Schematic of a TMDC basal plane. The leftmost and rightmost sides of the basal plane, marked by black dashed lines, show a TMDC of metal *A* (orange) and metal *B* (blue), respectively. The central region in between the two dashed lines shows the solid solution of the two metals forming a quasi-binary alloy. The dashed red circles denote S-vacancies, which are the sites where the $CO_2$RR and HER are found to occur preferentially. **b)** Free energy diagram of pristine $MoS_2$ (blue) and one with a S-vacancy (pink) on the basal plane at an applied potential of -0.11 V RHE, the equilibrium potential of the reaction. **c)** The top view schematics of the adsorption configuration are shown for the pristine $MoS_2$ (top) and $MoS_2$ with S-vacancy (bottom). **d)** Comparison of free energies between Ag (111)[2] and the five unary TMDCs considered in this study with a S-vacancy, also at an applied potential of -0.11V RHE.

### 3.2. Thermodynamic stability and sulfur vacancy formation in quasi-binary TMDCs

Before looking at adsorption energies, we first investigated the stability of the random solid solution and the feasibility of creating an S-vacancy as our first screening method for effective catalysts. While alloying opens the space of adsorption energies, there is a vast number of unique active sites in a random solid solution. However, it is known that the binding energy of reaction intermediates is largely influenced by the closest atoms to the active site[31]. Therefore, we can simplify the reaction site to a local environment of the 6 nearest transition metals closest to the S-vacancy, as shown in Fig. S1. Considering these six neighbors for a quasi-binary TMDC alloy results in a total combination of $2(2^6) = 128$ local environments. These 128 local environments can be further reduced to 20 symmetrically unique local environments, as illustrated in Fig. 2a. There are 10 local environments that correspond to the symmetrically unique patterns with respect to metal $A$ positions. As the 10 patterns could also be with respect to metal $B$, 20 local environments for every quasi-binary composition need to be accounted for. There are two pairs of local environments that have the same composition, denoted as $m$ and $x$. The difference in each pair reflects the absence or presence of a mirror symmetry, depicted with a dashed line in some of the configurations in Fig. 2a.

We evaluated the stability of the alloys using the enthalpy of mixing, $\Delta H_{mixing}$, calculated using DFT at 0 K using Eq. 1. A negative $\Delta H_{mixing}$ indicates that a quasi-binary alloy is thermodynamically miscible at 0 K and can form without phase segregation[25]. Of the ten alloys, only (Mo, W)$S_2$ and (Nb, Ta)$S_2$ exhibit negative $\Delta H_{mixing}$; all other binary TMDCs have positive values. These results are consistent with previous reports[16]. The ten TMDC alloys can therefore be grouped into three categories based on their mixing enthalpies. The first group consists of (Mo, W)$S_2$ and (Nb, Ta)$S_2$, which are thermodynamically stable and fully miscible at all temperatures. Notably, both alloys contain cations from the same group in the Periodic Table, either group V or group VI. The second group, consisting of (Mo, Nb)$S_2$, (Mo, Ta)$S_2$, (Nb, W)$S_2$, and (Ta, W)$S_2$, has slightly positive mixing enthalpies at 0 K (12.7 – 20.8 meV/f.u.) and contains mixtures of group V and group VI elements. The third group, consisting of (Mo, V)$S_2$, (Nb, V)$S_2$, (Ta, V)$S_2$, and (V, W)$S_2$, includes the vanadium-containing compositions, which show relatively high instability (16.2 – 42.4 meV/f.u.). Although V is also a group-V element, it is not miscible with Nb or Ta, distinguishing these alloys from the first group.

A similar grouping emerges from the S-vacancy formation energies ($\Delta E_{vac}$) across the different local environments of the ten binaries, as shown in Fig. 2c. The box-and-whisker plots compare both the spread and median vacancy formation energy for each alloy. At room temperature, the Boltzmann distribution indicates that the vacancy configurations having the lowest energy are expected to dominate. To illustrate the broader trend, the vacancy-energy distribution at 1000 K is shown in Fig. S2; however, the same overall behavior is evident whether the lowest energies or median energies are considered. Group 1 exhibits the narrowest distribution of vacancy formation energies, whereas Group 3 shows the widest distribution. Group 2 has relatively high vacancy formation energies, with medians above 2 eV, while Group 3 generally has lower values, with medians below 2 eV. Comparing the medians and interquartile ranges indicates that Group V elements generally lower sulfur-vacancy formation energies, whereas group VI elements increase them, as illustrated by the energy shift between (Mo, W)$S_2$ and (Nb, Ta)$S_2$.

These trends highlight a tradeoff between alloy stability and vacancy formation. While (Mo, W)$S_2$ and (Nb, Ta)$S_2$ are stabilized as random solid solutions, vanadium-containing binaries are expected to have higher vacancy concentrations and therefore may enhance $CO_2$RR activity. However, V-containing alloys also exhibit substantially larger energy spreads in vacancy formation energy than the other compositions. In contrast, (Nb, Ta)$S_2$ shows remarkable site uniformity: its vacancy formation energies vary by only 0.22 eV/f.u. around a median of 1.88 eV/f.u., compared with a broader 2.02 eV/f.u. spread around a median of 1.80 eV/f.u. for the V-containing alloys. This low variability is advantageous for catalysis because a narrow energy distribution is expected to result in a more homogeneous population of active sites, leading to more predictable kinetics and reducing the likelihood of local environments that favor undesired side reactions. We further note that the vacancy formation energies have been calculated using sulfur-rich chemical potentials. For intermediate or sulfur-poor chemical potentials, these values will be lower, as shown in SI Fig. S3, resulting in a higher concentration of vacancies at room temperature.

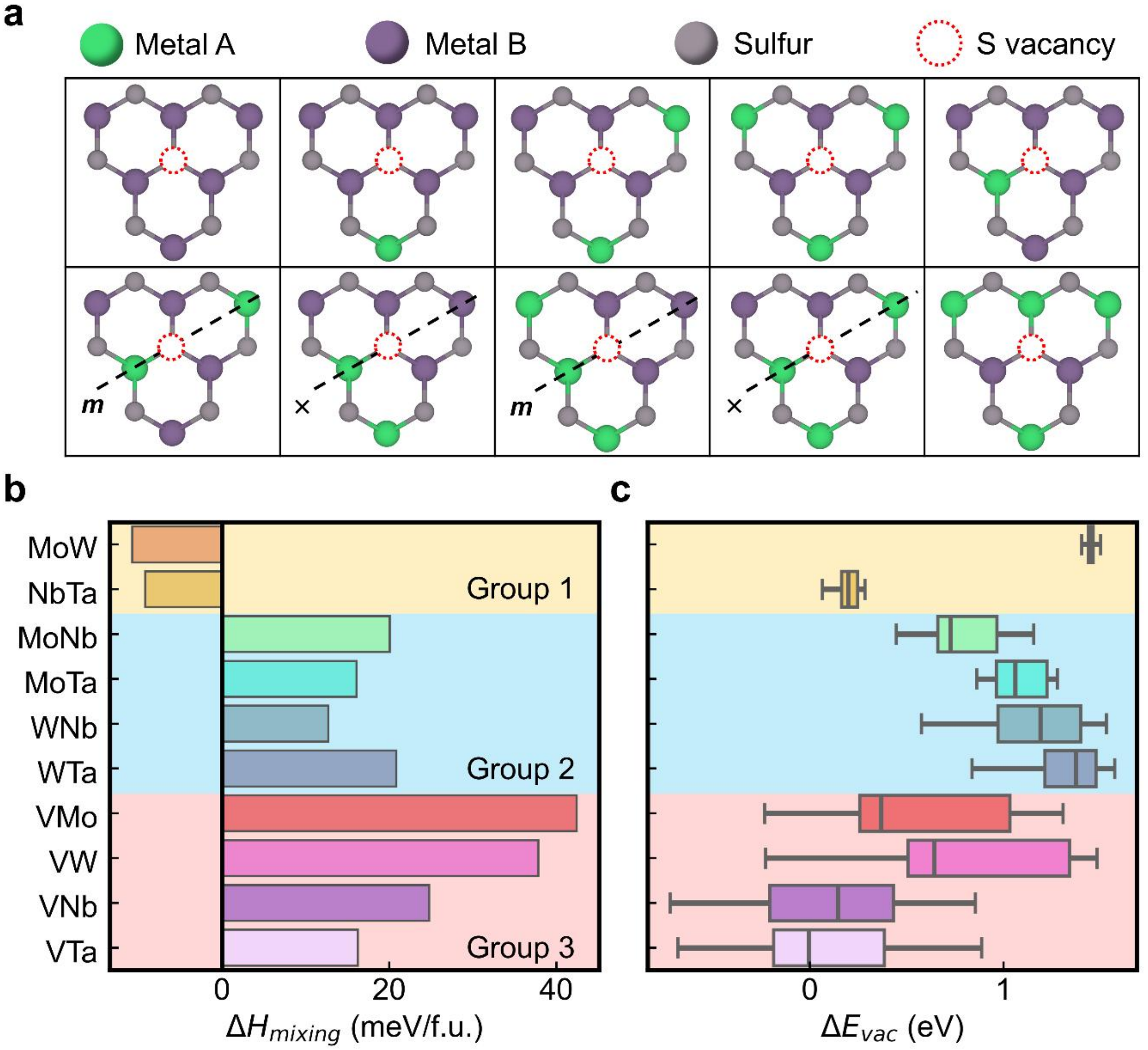

**Figure 2: a)** The ten symmetrically unique local environments considering the nearest and next-nearest cations around an S vacancy in a quasi-binary TMDC alloy. Local environments with an *m* or *x*, respectively, show a mirror symmetry, or lack thereof, along the dashed line. **b)** The mixing enthalpy of the ten binary TMDCs at 0 K. **c)** Box and whisker plots of the formation energy of an S-vacancy for the TMDC alloys at all unique local environments. Binaries are grouped by pure group **V/VI** alloys (yellow), mixed group alloys without vanadium (blue), and vanadium containing alloys (red).

### 3.3. Adsorption-energy trends reveal an inverted *d*-band scaling relation

To conduct further screening for efficient $CO_2$RR catalysts amongst the quasi-binary TMDCs, the binding energies of the intermediate steps to the reaction, COOH* and CO*, were calculated. Across S-vacancy sites with different local coordination environments, Group 1 alloys exhibit a narrow adsorption-energy distribution, whereas Group 3 alloys show the broadest spread, as shown in Fig. 3a- b. This is consistent with the spread in sulfur vacancy formation energies seen previously. In general, Group 2 binaries show strong binding with mostly negative binding energies with both COOH* and CO*, whereas Group 3 binaries have weak binding to COOH*, with most energies being above 0 eV, and relatively moderate binding to CO*, being close to 0 eV. There are notable differences in the binding energies of $(Mo, W)S_2$ and $(Nb, Ta)S_2$, both of which belong to Group 1. The CO* binding energies for both are very moderate, but $(Mo, W)S_2$ exhibits the weakest COOH* binding energies across all ten binaries (1.18 eV), while $(Nb, Ta)S_2$ retains near zero binding energies.

To connect the trends in adsorption energies to the electronic structure, we employ the *d*-band center model as a descriptor of the binding energy strength[26]. The projected density of states (pDOS) of transition metals participating in adsorption can be used as an indicator for the binding energy between the metals and the adsorbate. Contrary to the standard *d*-band model, where a lower *d*-band center weakens binding on transition metal surfaces[32], TMDCs display an inverted trend. In these 2D TMDC alloys, a downshifted *d*-band center leads to more negative binding energies between the adsorbate and the catalyst[27,28]. A schematic showing the differences between the standard *d*-band model and the inverted trend can be seen in SI Fig. S4. This trend is evident across several binaries. For example, Group 3 vanadium-containing alloys exhibit relatively higher *d*-band centers for COOH* than the other binaries, as shown in the bottom-panel of Fig. 3a, consistent with their weaker COOH* binding with positive binding energies, as shown in the top-panel in Fig. 3a. This trend persists across the metallic quasi-binary TMDCs studied here, but fails for $(Mo, W)S_2$. The semiconducting nature of $(Mo, W)S_2$, which is unique among the ten binaries tested, disrupts the electronic coupling required for the inverted *d*-band model to hold, marking a clear boundary for the model's applicability. As noted above, both the intermediates bind to $(Nb, Ta)S_2$ with moderate, near-zero adsorption energies, placing its limiting potential close to zero and indicating high intrinsic $CO_2RR$ activity. By the same criterion, $(Nb, V)S_2$ may also appear active. However, screening must account for both the magnitude and the distribution of binding energies across local environments. For example, $(Nb, V)S_2$ spans a broad adsorption-energy range from -0.155 to 0.815 eV, indicating strong site-to-site variability; where some sites are expected to bind strongly with the intermediates, which would hinder the subsequent reaction steps. In contrast, (Nb, Ta)$S_2$ combines near-optimal adsorption energies for both COOH* and CO* with a narrow energy distribution across local environments. This combination makes (Nb, Ta)$S_2$ the strongest candidate among the quasi-binary TMDCs screened here for an intrinsically active basal-plane $CO_2$RR catalyst.

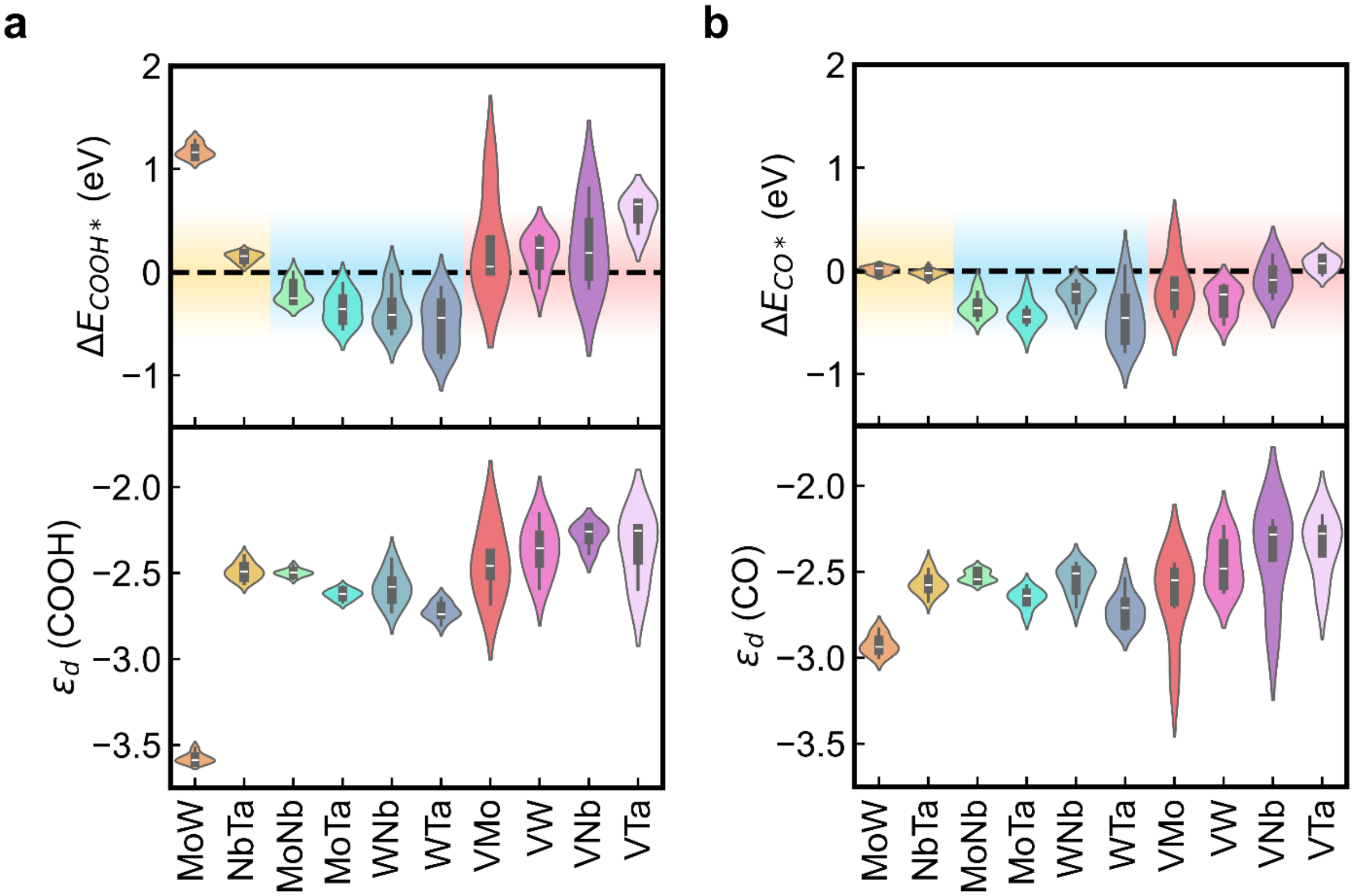

**Figure 3: a-b)** Violin distribution plots of CO*/COOH* binding energies at an applied potential of 0 V *vs.* RHE, and the *d*-band center of the ten alloys for varying local environments with the same grouping as Fig. 2.

### 3.4. Local site-coordination controls $CO_2$RR activity and HER selectivity

Besides activity, a viable $CO_2$RR catalyst also must exhibit high selectivity against competing reactions. A challenge of $CO_2$RR catalysts is their inability to limit the hydrogen evolution reaction (HER)[10,33].

Evaluating selectivity requires a comparative analysis of the binding energies of the HER intermediates. Following Sabatier's principal[34], optimal catalysts require a balance where intermediates do not bind too strongly or weakly. We evaluated this trade off by constructing volcano plots to identify candidates with minimized limiting potentials ($U_L$). Fig. 4a shows the $CO_2$RR volcano plot, where the red (left) slope represents sites limited by CO desorption, while the blue (right) slope signifies limitation by COOH* formation. All the Group 2 quasi-binary TMDC's are limited by the desorption of CO, and many Group 3 TMDC's are limited by the adsorption of $CO_2$ to COOH*. The trend reflects vanadium's tendency to weaken the binding energy between the COOH adsorbate and the catalyst, as discussed in Fig. 3a. We can also see that (Nb, Ta)$S_2$ largely clusters at the optimum of the volcano, indicating its low $U_L$, regardless of the local environment. It can also be seen that (Mo, W)$S_2$ (orange) exhibits the largest $U_L$, due to the weak binding energy of COOH*.

Beyond $CO_2$RR activity, we have also evaluated the selectivity of these TMDC alloys by examining the energetics of the competing HER pathway, which proceeds through $H^+$ adsorption to form H* followed by $H_2$ desorption. The resulting Sabatier volcano plot for HER is shown in Fig. 4b. The left side of the plot shows catalysts that are limited by the desorption of ½$H_2$, and the right side shows those limited by the adsorption of $H^+$ to H*. The volcano's characteristic linearity arises from the single intermediate (H*) involved in the HER pathway. Much like the volcano in Fig. 4a, TMDC alloys that lie at the proximity of the peak are strong catalysts for this reaction. However, for the alloy to be optimal for $CO_2$RR, we would want it to lie at the peak of the $CO_2$RR volcano and away from the peak of the HER volcano. We find that all Group 2 TMDC alloys are limited by strong H binding, whereas Group 3 can be limited by either step, depending on the composition and local environment, as shown in Fig. 4b. Looking at Group 1, both the TMDC alloys cluster at the minimum of the HER volcano, with (Nb, Ta)$S_2$ being slightly more spread away from the peak compared to (Mo, W)$S_2$.

Since (Nb, Ta)$S_2$—the strongest candidate for the $CO_2$RR—lies near the peak of both volcanos, a direct comparative analysis of their limiting potentials is required to quantify selectivity. Fig. 4c shows the $U_L$ of the $CO_2$RR with the difference between the $U_L$ from the $CO_2$RR and the HER, which we denote as $\Delta U_L$. This representation captures both selectivity and activity. Negative $\Delta U_L$ values indicate preference for $CO_2$RR, while $U_L$ values closer to zero correspond to higher $CO_2$RR activity. Many Group 2 alloys are selective against HER and show an average $CO_2$RR $U_L$ of ~0.4 V. Conversely, many Group 3 alloys are not selective against HER, and tend to also have a higher average $CO_2$RR $U_L$ of ~0.6 V. There are a few Group 3 alloys that exhibit selectivity, and fewer still that show low $CO_2$RR $U_L$ such as (Ta, V)$S_2$ and (Nb, V)$S_2$. While these could be seen as good candidates for a $CO_2$RR catalyst, Group 3 exhibits high variability in activity at different reaction sites. For instance, looking at (Nb, V)$S_2$ in the inset of Fig. 4c, there is a local environment in the alloy that is both selective and active, having a $CO_2$RR $U_L$ of ~0.05 eV; however, looking at the entire space, it can also be seen that there are points that not only have high $CO_2$RR $U_L$, but also show a preference for the HER indicated by positive $\Delta U_L$ values. That variability also extends to the other alloys in Group 3, indicating that they may not be the most reliable candidates for the $CO_2$RR reaction.

Notably, we can see the disparity between the alloys in Group 1. (Mo, W)$S_2$ exhibits the lowest selectivity towards the $CO_2$RR, showing a strong preference for the HER, with the high $U_L$ for $CO_2$RR indicating low activity. Thus, although $MoS_2$ and $WS_2$ have been reported to be effective catalysts for $CO_2$RR when their edge sites are activated, for instance in ionic liquids[5], the (Mo, W)$S_2$ basal plane does not exhibit comparable activity or selectivity. However, we find the other catalyst in the group, (Nb, Ta)$S_2$, to be an active and selective catalyst for $CO_2$RR. Most local environments exhibit near-zero $CO_2$RR $U_L$ values, averaging ~0.05 V, with little sensitivity to local atomic configurations, as shown in the inset of Fig. 4c. Many local environments are also selective for $CO_2$RR, with the non-selective local environments showing only a slight preference for HER.

To further analyze the disparity in the two Group 1 TMDC alloys, we show the binding configurations in the local environment of active and inactive catalysts for both reactions in Group 1 in Fig. 4d. From the $CO_2$RR column, we can see the difference in the oxygen-TM binding configurations between the two catalysts during COOH* adsorption, the reaction step that creates the energy disparity within Group 1. In (Nb, Ta)$S_2$, the $sp^2$-hybridized oxygen atom coordinates with two neighboring transition metals, forming a stable bidentate geometry. Furthermore, there is a tilting effect with the C atom towards the third transition metal in the first shell. Conversely, a different binding behavior for COOH* can be observed for (Mo, W)$S_2$. While there is still a similar C tilting effect, instead of the bidentate bonding seen for the $sp^2$ oxygen atom in (Nb, Ta)$S_2$, we see a monodentate behavior, where one of the transition metals in the first shell does not coordinate to oxygen or has carbon tilting toward it. These binding configurations are representative of the 16 (Mo, W)$S_2$ and (Nb, Ta)$S_2$ local environments tested (as shown in SI Fig. S5 and S6). Thus, we can see a correlation with the binding configuration and the energies of COOH* adsorption in the Group 1 alloys. Since (Mo, W)$S_2$ experiences monodentate oxygen bonding and very weak COOH* adsorption energy (1.26 eV), it can be concluded that this binding configuration leads to high adsorption energy, as opposed to the bidentate oxygen bonding seen in (Nb, Ta)$S_2$.

The most active local composition for $CO_2$RR is 33% of Nb and 66% Ta in the 1$^{st}$ and 2$^{nd}$ shell, despite Ta showing the best activity in its unary form. A similar bonding analysis can be done for the difference in the relative binding energies in HER. Looking at the bonding behavior for HER, it does not appear that either catalyst shows a direct TM-adsorbate bond like COOH*. Instead, the difference in the binding energies can be seen in the amount of hydrogen displacement off the central sulfur atom and towards the first shell transition metals, shown by the dashed red circles in Fig 4d. (Mo, W)$S_2$ causes the hydrogen atom to move towards two of the transition metals more than the (Nb, Ta)$S_2$ site, which stays mostly at the center of the local environment. The tilting in (Mo, W)$S_2$ would therefore lead to more bonding interactions between the H and two of the transition metals in the first shell, enhancing the binding. Notably, the local composition of (Nb, Ta)$S_2$ that gives the highest selectivity is one that is mostly comprised of Ta (5 out of the 6 cations). The optimal local composition for selectivity in (Nb, Ta)$S_2$ brings back the selectivity issue from Fig. 4c. To further investigate how composition affects the selectivity of (Nb, Ta)$S_2$, the average activity is compared for both $CO_2$RR and HER in (Nb, Ta)$S_2$ as a function of the Ta content in the first shell, seen in Fig 4e. The $CO_2$RR $U_L$ values change negligibly until Ta makes up all three atoms in the first shell, indicating that the presence of Ta in the first shell has little effect on $U_L$. On the other hand, the HER goes from having an average $U_L$ of 0.03 V to 0.06 V by just introducing a single Ta atom in the first shell, with the trend increasing as more Ta is incorporated. It can therefore be concluded that the presence—or absence—of Ta in (Nb, Ta)$S_2$ is one of the primary factors for the selectivity differences seen in Fig. 4c. Even with these differences in selectivity, we conclude that of the 10 catalysts we have investigated, (Nb, Ta)$S_2$ shows the most promise as an active and selective catalyst for $CO_2$RR.

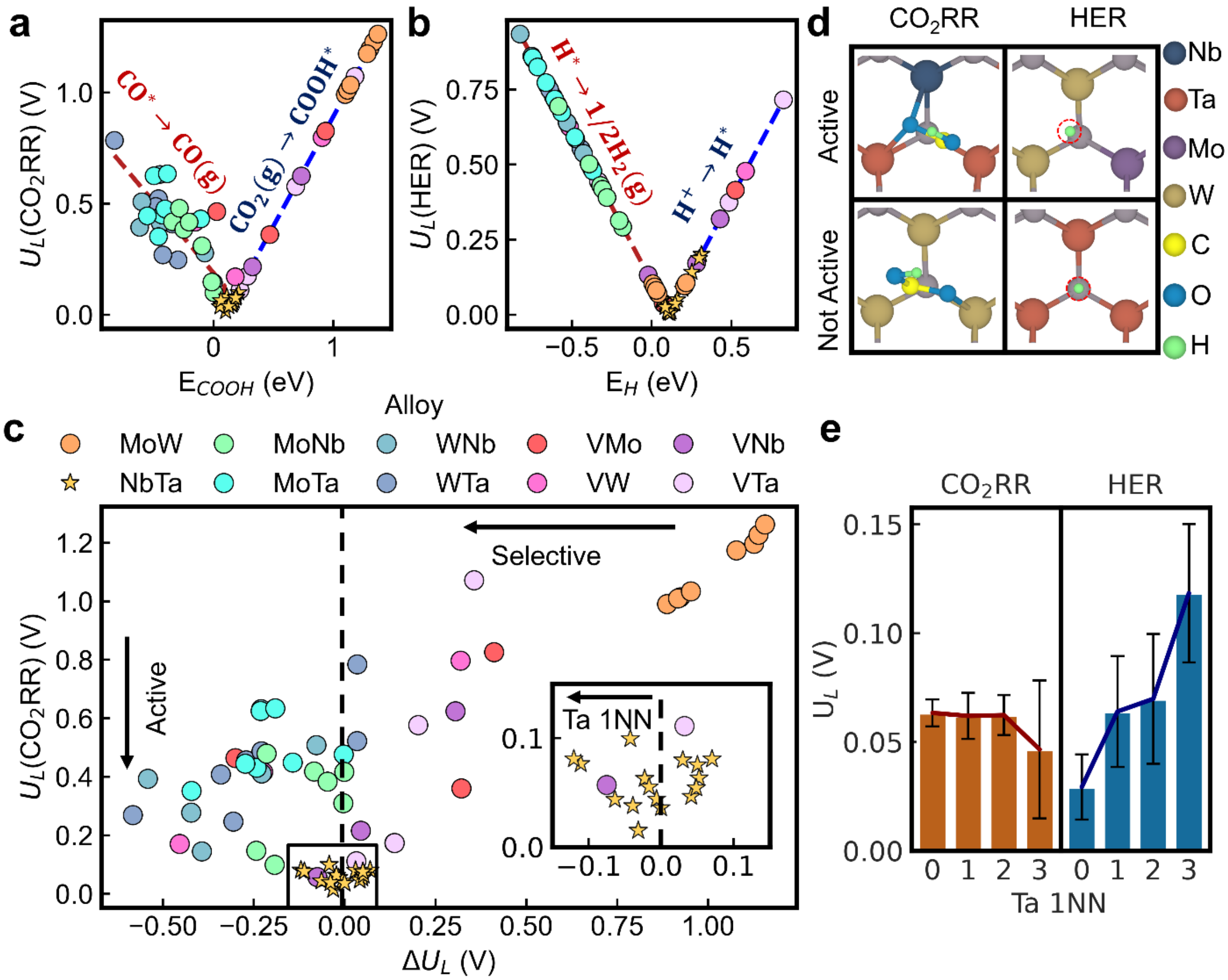

**Figure 4: a)** The Sabatier volcano for $CO_2$RR at $U = -0.11$V. Most alloy environments that align on the left (red) line are limited by the desorption of CO. Environments that align on the right (blue) line are limited by the adsorption of $CO_2$ to COOH*. **b)** The Sabatier volcano for HER at $U = -0.11$V. Alloy environments that align on the left (red) are limited by the desorption of H*. Environments that align on the right (blue) line are limited by the adsorption of $H^+$ to H*. **c)** The selectivity of the alloy local environments against HER. Local environments in the green region have optimal $U_L$ for $CO_2$RR. Local

environments to the left of the vertical dashed line have selectivity against HER. The inset shows the local environments at the selectivity line with low $CO_2$RR $U_L$. **e)** Binding configurations at the limiting intermediate step for active and inactive catalysts in group 1. **f)** Trend of $U_L$ for $CO_2$RR and HER in (Nb, Ta)$S_2$ local environments, grouped by Ta composition in the 1$^{st}$ nearest neighbor shell surrounding the sulfur vacancy.

### 3.5. Electronic origin of intermediate adsorption via pDOS and COHP analysis

As stated previously, the *d*-band center is a qualitative descriptor of binding strength for intermediate steps[32]. To better resolve the bonding mechanism, we performed Crystal Orbital Hamilton Population (COHP) analysis, which quantifies bonding and antibonding interactions between adsorbates and the atoms they bind to. Specifically, we performed COHP analysis to better understand the mechanism behind the binding energy of the CO* intermediate step. Since the CO* step in $CO_2$RR does not rely on $H^+/e^-$ transfer, unlike the COOH* intermediate step, it is primarily influenced by the bonding/antibonding contributions, as opposed to the applied potential in the reaction cell. Fig. 5(a-c) shows the partial density of states (pDOS) of TM *d*-states and CO *p*-states, and -COHP of (Mo, Ta)$S_2$, (Nb, Ta)$S_2$, and (Ta, V)$S_2$ active sites, centered on the S-vacancy, which represent the three groups, all of which contain Ta. Ta is a useful reference because it shows the highest intrinsic activity among the unary TMDCs, yet only modestly lowers the $CO_2$RR $U_L$ in (Nb, Ta)$S_2$. To assess the adsorbate-catalyst hybridization, we examine the energetic overlap between the CO *p*-states and the transition metal *d*-states, across the three configurations. Fig. 5a shows a local environment for (Mo, Ta)$S_2$, a composition from Group 2, which binds strongly to CO*. In the pDOS, the peak, the strongest energetic overlap, close to the Fermi energy between the transition metals and CO is ~ –0.9 eV. By looking at the binding configuration, despite having two Ta in the nearest neighbors, the CO atom has preferential coordination to the Mo atom, suggesting that for this system, Mo strongly contributes to the hybridized peaks and therefore, strong CO binding. Conversely, in Fig. 5b, we see that the transition metals in the (Nb, Ta)$S_2$ environment, a representative of Group 1, have a peak at ~ −0.5 eV. In this case, once again, CO coordinates to the Nb and not towards the two Ta. This would suggest that the presence of Ta, despite the lack of direct coordination, modulates the binding strength. This is further exemplified in Fig. 5c, which shows (Ta, V)$S_2$ from Group 3 and has the weakest binding energy of the three. Here, CO coordinates directly with Ta, yielding minimal TM-adsorbate peak alignment and sparse CO *p*-states near the Fermi energy.

To further explore the binding behavior, the sum of the -COHP of the C atom and its nearest neighbors is shown in Fig. 5a-c (middle panel). In -COHP, positive states indicate bonding behavior, whereas negative states indicate anti-bonding behavior[35]. States with a value of zero indicate that there is no bonding between the C atom and the catalyst surface. In all the three local environments, both bonding and anti-bonding states are observed below the Fermi energy. Looking first at (Mo, Ta)$S_2$, there is a bonding peak at ~ –0.9 eV, corresponding to the hybridized peaks in the pDOS. As we move away from the Fermi energy, the bonding states decrease and become anti-bonding states at ~ –3.8 eV. From –6 to –3.8 eV, there appear to be more anti-bonding states than bonding states at the tail of the bonding peak (–3.8 to +1.5 eV). The -COHP also provides the basis for the inverse *d*-band center trend observed in Fig. 3. In transition metal catalysts, the *d*-band center connects the binding strength to the population of anti-bonding states below the Fermi energy[32]. However, our analysis shows that the bonding states are primarily above the Fermi energy, and thus unfilled. The presence of the unfilled bonding states means that a *d*-band center closer to the Fermi energy would indicate more bonding states being unpopulated. The unpopulated bonding states create a deviation from the classical *d*-band model, and explains why downshifted *d*-band center leads to a stronger binding energy. Furthermore, we can use the integrated COHP (ICOHP) to measure the amount of bonding states compared to anti-bonding states present up to the Fermi energy. The more bonding states there are, the higher the ICOHP will be. A higher ICOHP then correlates to a stronger, or more negative, adsorption energy[33]. It can be seen that (Mo, Ta)$S_2$ has an ICOHP of 7.778. (Nb, Ta)$S_2$, by comparison, has an ICOHP of 5. 661. Plots of the ICOHP can be found in SI Fig. S7. Since (Mo, Ta)$S_2$ has a stronger adsorption energy than (Nb, Ta)$S_2$, we expect the ICOHP to also be greater. We can also see in the -COHP plot of (Nb, Ta)$S_2$ that the bonding peak at ~–0.5 eV is less intense than (Mo, Ta)$S_2$, corresponding to the weaker adsorption energy seen in (Nb, Ta)$S_2$. Furthermore, we expect (Ta, V)$S_2$ to have the lowest value for ICOHP as it has the weakest adsorption energy of the three local environments, which holds true, as (Ta, V)$S_2$ has an ICOHP of 4.911. This is further confirmed in the -COHP, where the bonding peak nearest to the Fermi level is the least intense of the three environments. Analysis of both the pDOS and -COHP provides deeper insight into adsorption behavior across the three catalyst groups and explains why (Nb, Ta)$S_2$ exhibits the most moderate binding energy.

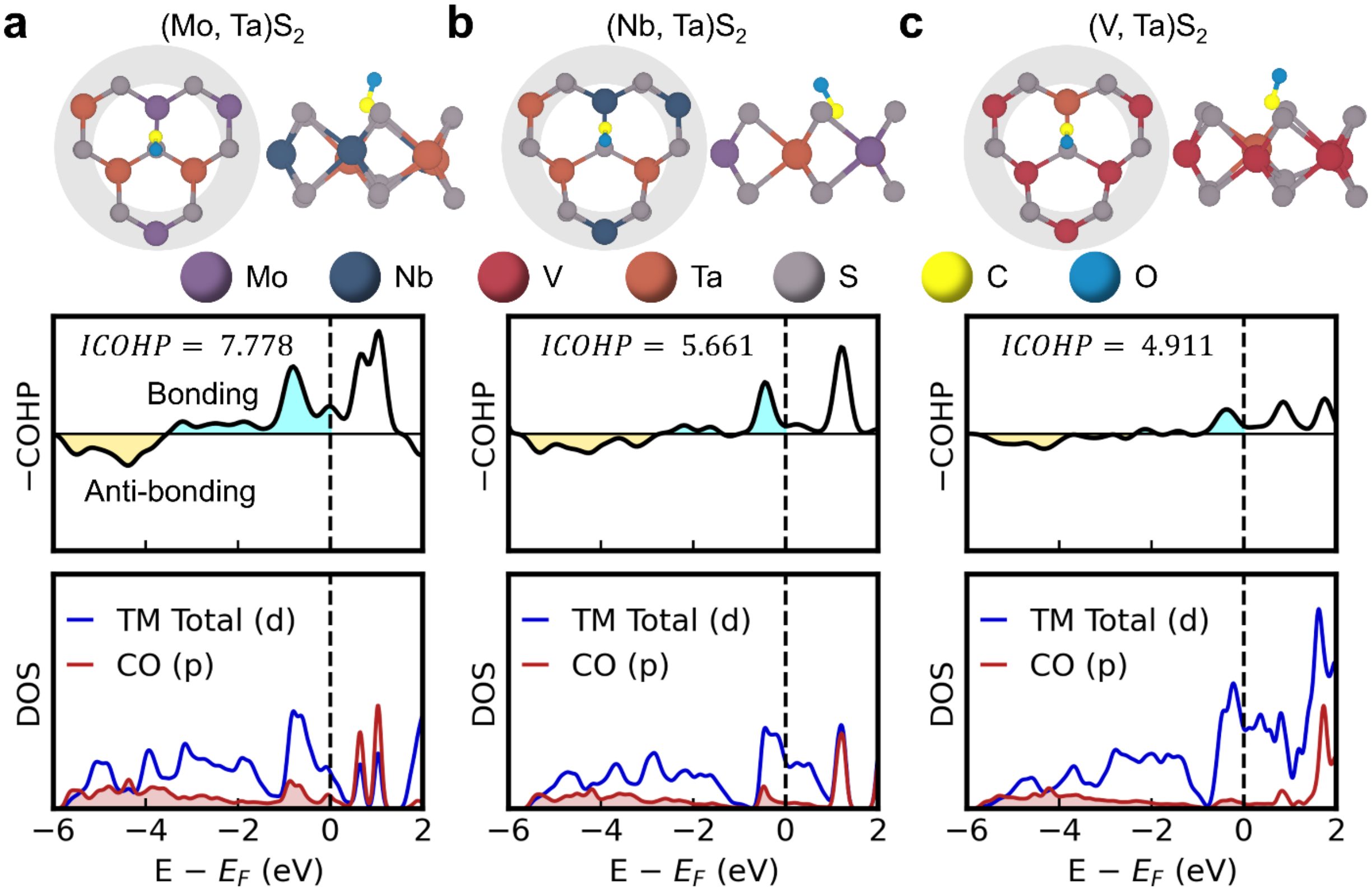

**Figure 5:** The -COHP and pDOS of three representative local environments from each of the groups, with the bonding configuration shown above for **a)** (Mo, Ta)$S_2$ **b)** (Nb, Ta)$S_2$ **c)** and (V, Ta)$S_2$. Regions in the -COHP shaded blue are bonding states. Regions shaded in yellow are anti-bonding states. The density of CO *p*-states and TM *d*-states are shown using red and blue lines, respectively, in the bottom panels. The vertical dashed lines represent the Fermi energy.

## 4. Conclusions

Through DFT calculations, we have established a dual-functional strategy to both activate the TMDC basal plane, while systematically tuning $CO_2$RR intermediate binding energies. Sulfur vacancies serve to thermodynamically activate the basal plane by creating local active sites, whereas alloying TMDCs as solid-solutions broadens the accessible range of binding energies. By integrating these two strategies, we identify (Nb, Ta)$S_2$ as a promising catalyst exhibiting near optimal binding energies for both the $CO_2$RR intermediates. We observe trends such as alloying Mo or W with either Nb or Ta yields catalysts with strong binding in both steps, while vanadium-containing alloys show weak COOH* binding.

Binding energy trends correlate with the *d*-band center model, though notably, the TMDC alloys studied here exhibit an inverted relationship compared to traditional transition metal surfaces[32]. Selectivity evaluations against the competing HER revealed that (Mo, W)$S_2$ is hindered by thermodynamic preference for hydrogen evolution, coupled with suboptimal $CO_2$RR activity. Conversely, (Nb, Ta)$S_2$ shows lower $U_L$ for $CO_2$RR and a higher $U_L$ for HER. Furthermore, our analysis revealed that the presence of Ta in the active site results in an increase in $U_L$ values for HER. Electronic structure analysis through COHP reveals that the binding disparity is rooted in the distribution of bonding and anti-bonding states, particularly the modulation of unoccupied bonding states above the Fermi energy.

Overall, this work establishes a predictive framework for designing TMDC alloy catalysts by linking composition, local coordination, defect formation, and electronic structure to catalytic activity and selectivity. Our results identify narrow distributions of site-specific energetics as a key design criterion and provide a bonding-based framework for systems where the conventional *d*-band model fails. Extending this approach to multi-principal-element TMDCs could enable basal-plane catalysts that surpass (Nb, Ta)$S_2$ in both $CO_2$RR activity and selectivity.

**Data Availability:** The data that supports the results of this study are openly available in the repository[37] "*Activating Basal Planes in Transition Metal Dichalcogenides for $CO_2$ Reduction to CO through Alloying*" at 10.5281/zenodo.22115193.

**Acknowledgements**

E.M.M was supported by the Ann W. and Spencer T. Olin-Chancellor's fellowship from Washington University. P.O., G.Y.J., R.M. were supported the National Science Foundation (NSF) through grant nos. DMR-2145797 (P.O., G.Y.J., R.M.) and DMR-2532378 (R.M.). This work used computational resources through allocation DMR160007 from the Advanced Cyberinfrastructure Coordination Ecosystem: Services & Support (ACCESS) program, which is supported by NSF grants #2138259, #2138286, #2138307, #2137603, and #2138296.

**Activating Basal Planes in Transition Metal Dichalcogenides for $CO_2$ Reduction to CO Through Quasi-binary Alloying**

**Eric Montufar-Morales[a], Daniel Rinder[b], Pravan Omprakash[a], Rohan Mishra[a,c*] and Gwan Yeong Jung[c,d*]**

[a] Institute of Materials Science & Engineering, Washington University in St. Louis, St. Louis, MO, USA

[b] Department of Computer Science & Engineering, Washington University in St. Louis, St. Louis, MO, USA

[c] Department of Mechanical Engineering & Materials Science, Washington University in St. Louis, St. Louis, MO, USA

[d] Department of Chemistry, Incheon National University, Incheon, 22012, Republic of Korea

Email: e.b.montufar-morales@wustl.edu; rmishra@wustl.edu; gyjung@inu.ac.kr

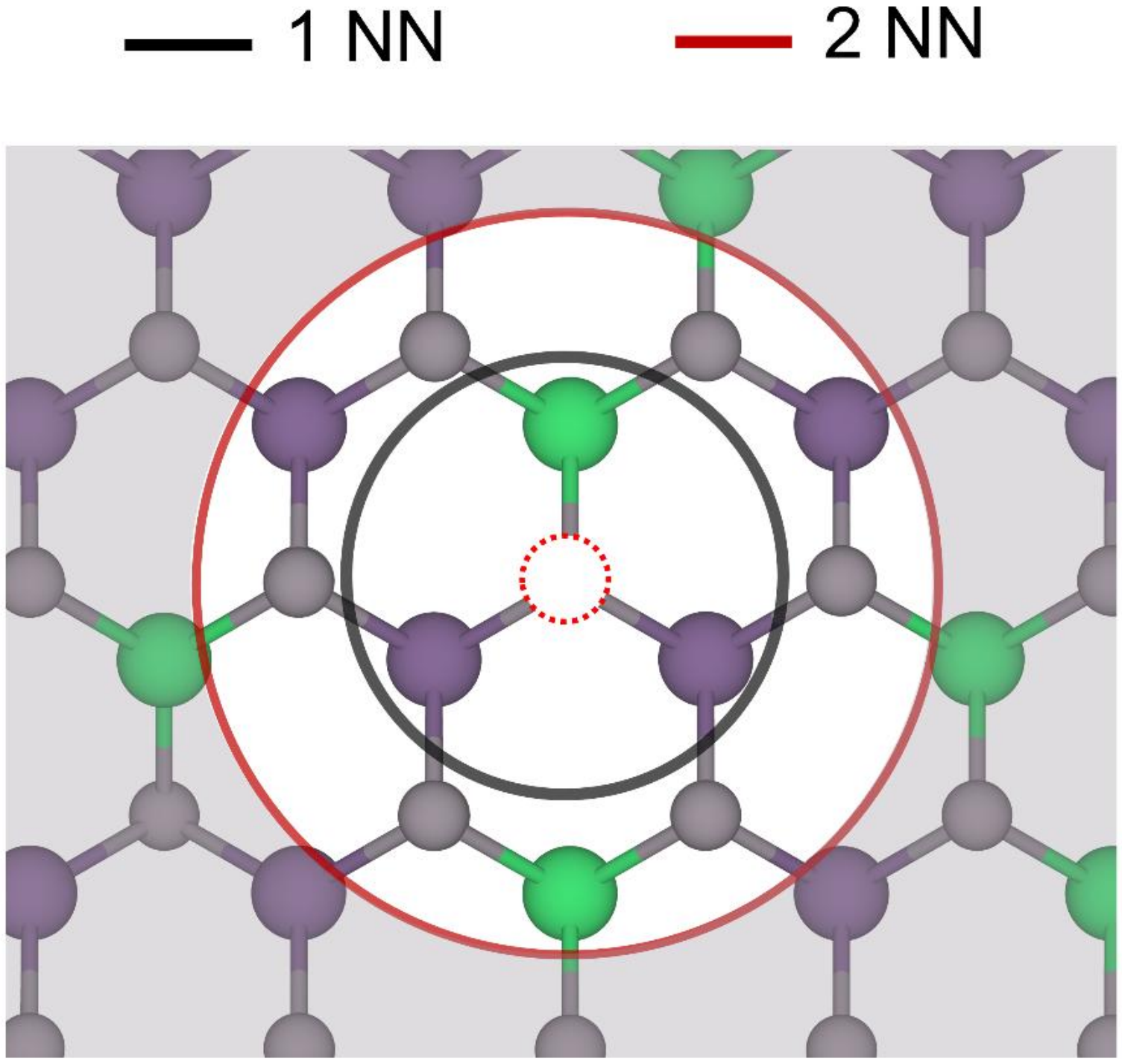

Figure S1. 1H binary TMDC viewed along the [001] direction. Green and purple atoms represent two different transition metals. Yellow atoms represent sulfur atoms. The Dotted red circle represents the location of a basal plane vacancy. Atoms enclosed by the black circle are the 1$^{st}$ nearest neighbors to the sulfur vacancy. Atoms enclosed by the red circle are the 2$^{nd}$ nearest neighbors to the sulfur vacancy.

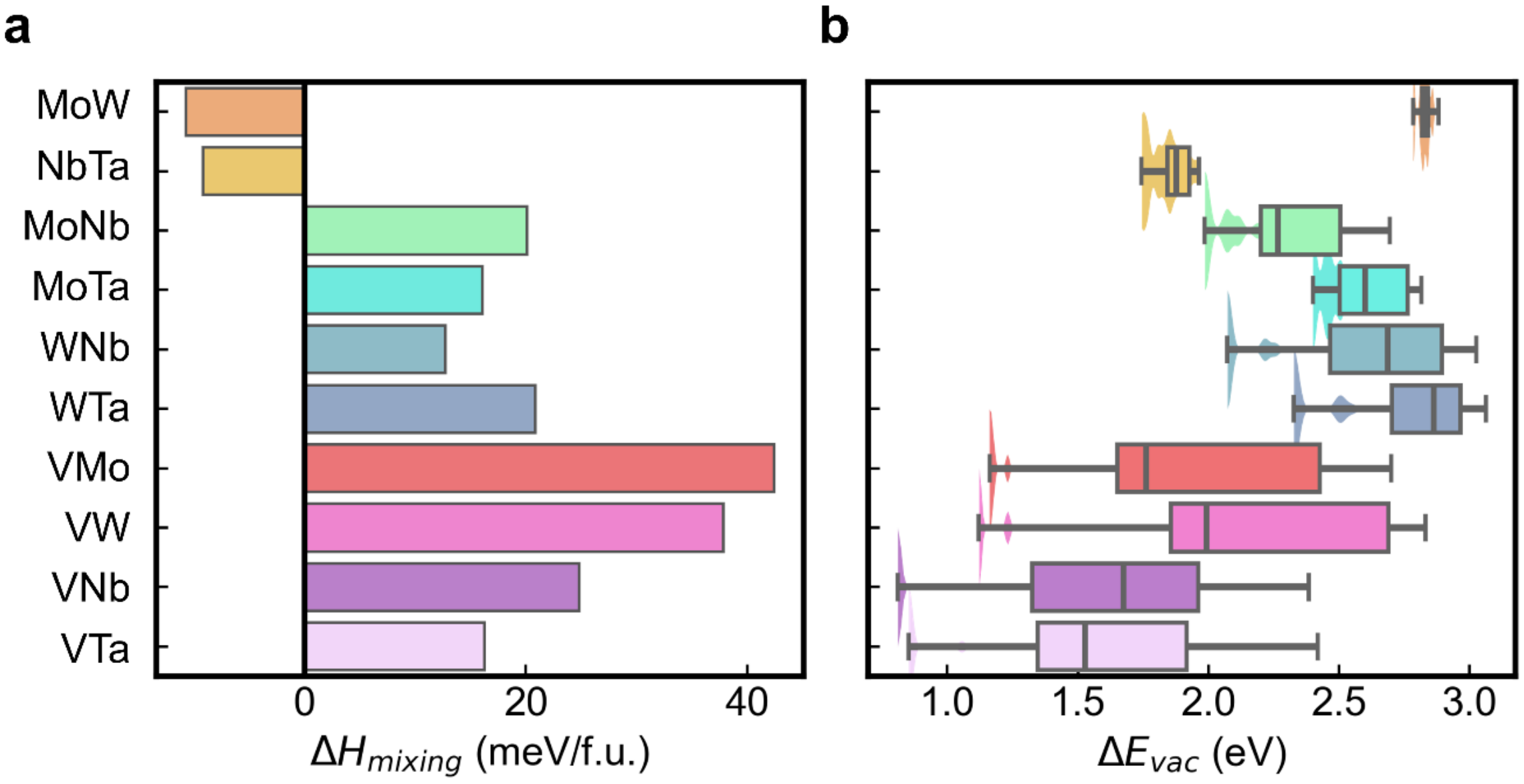

Figure S2. **a)** Enthalpy of Mixing for all 10 quasi-binary TMDCs at a 50:50 composition for the transition metals. **b)** Box and whisker plots of the Sulfur Vacancy Formation energy across various local environments. Underlaid are violin plots of the Boltzmann distribution of energies at 1000 K.

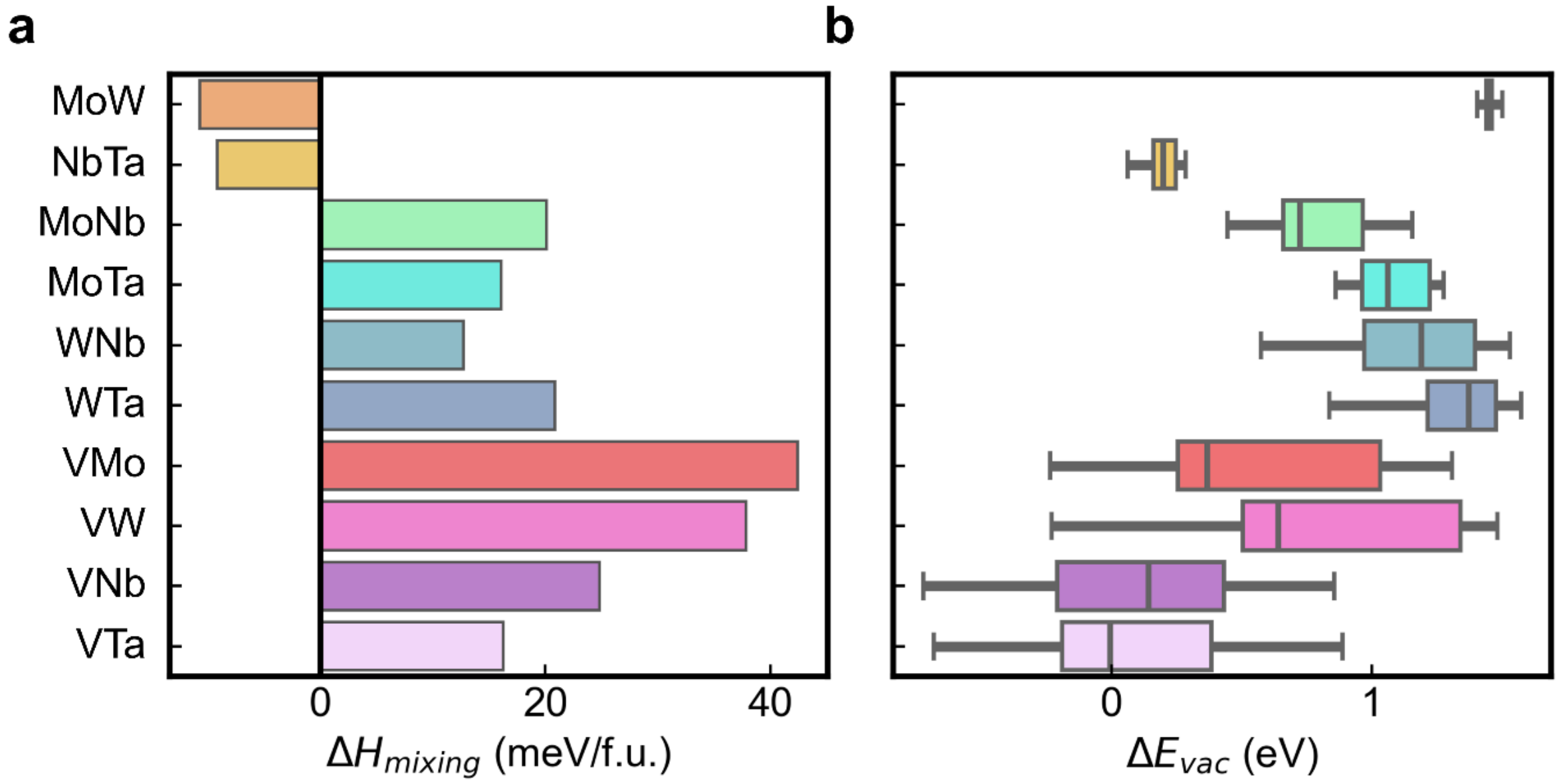

Figure S3. **a)** Enthalpy of Mixing for all 10 quasi-binary TMDCs at a 50:50 composition for the transition metals. **b)** Box and whisker plots of the Sulfur Vacancy Formation energy across various local environments under a Sulfur poor, transition metal rich environment

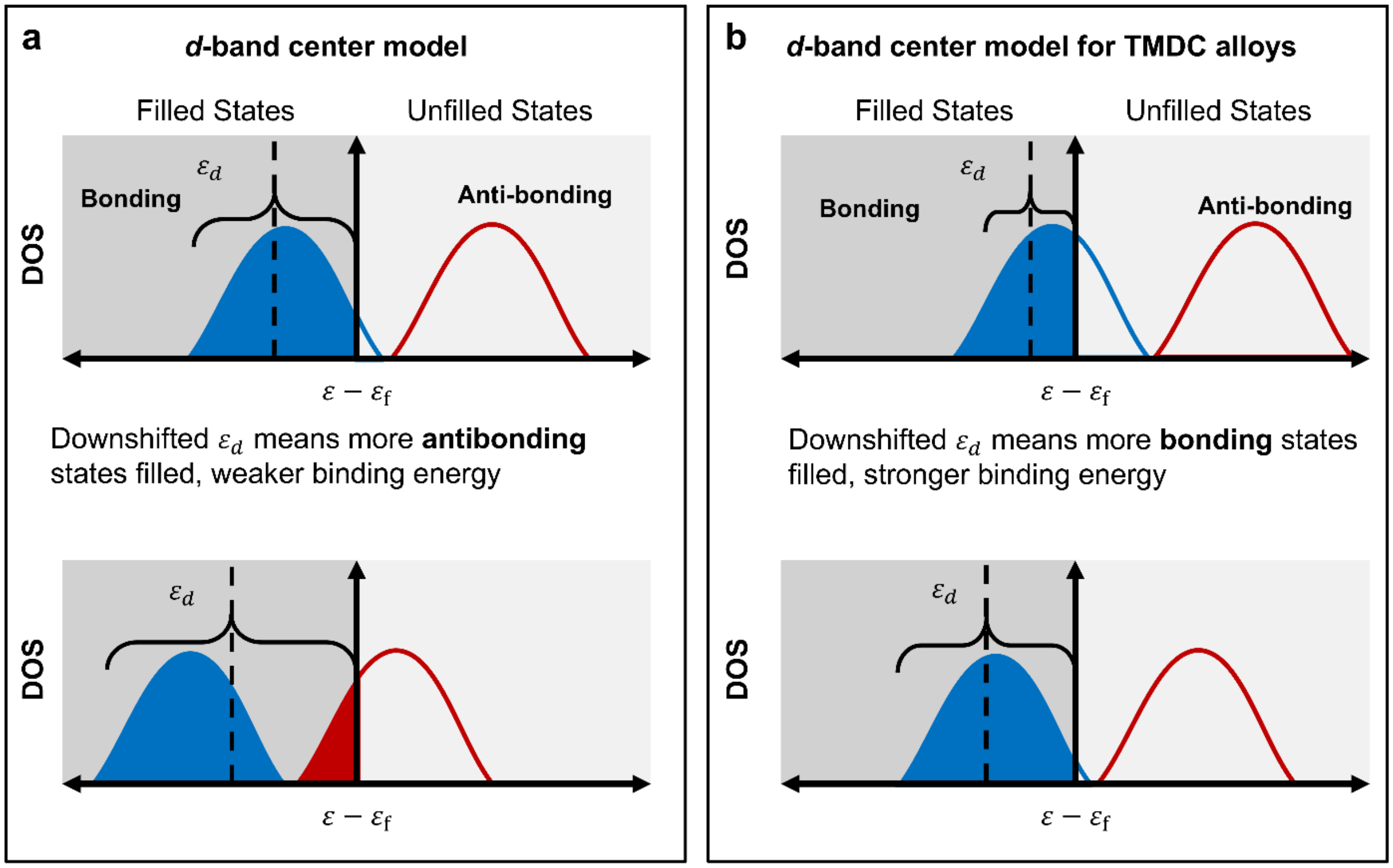

Figure S4. Difference between the *d*-band theory model in **a)** metals and **b)** transition metal dichalcogenides, distinguished by the different orbital filling behaviors.

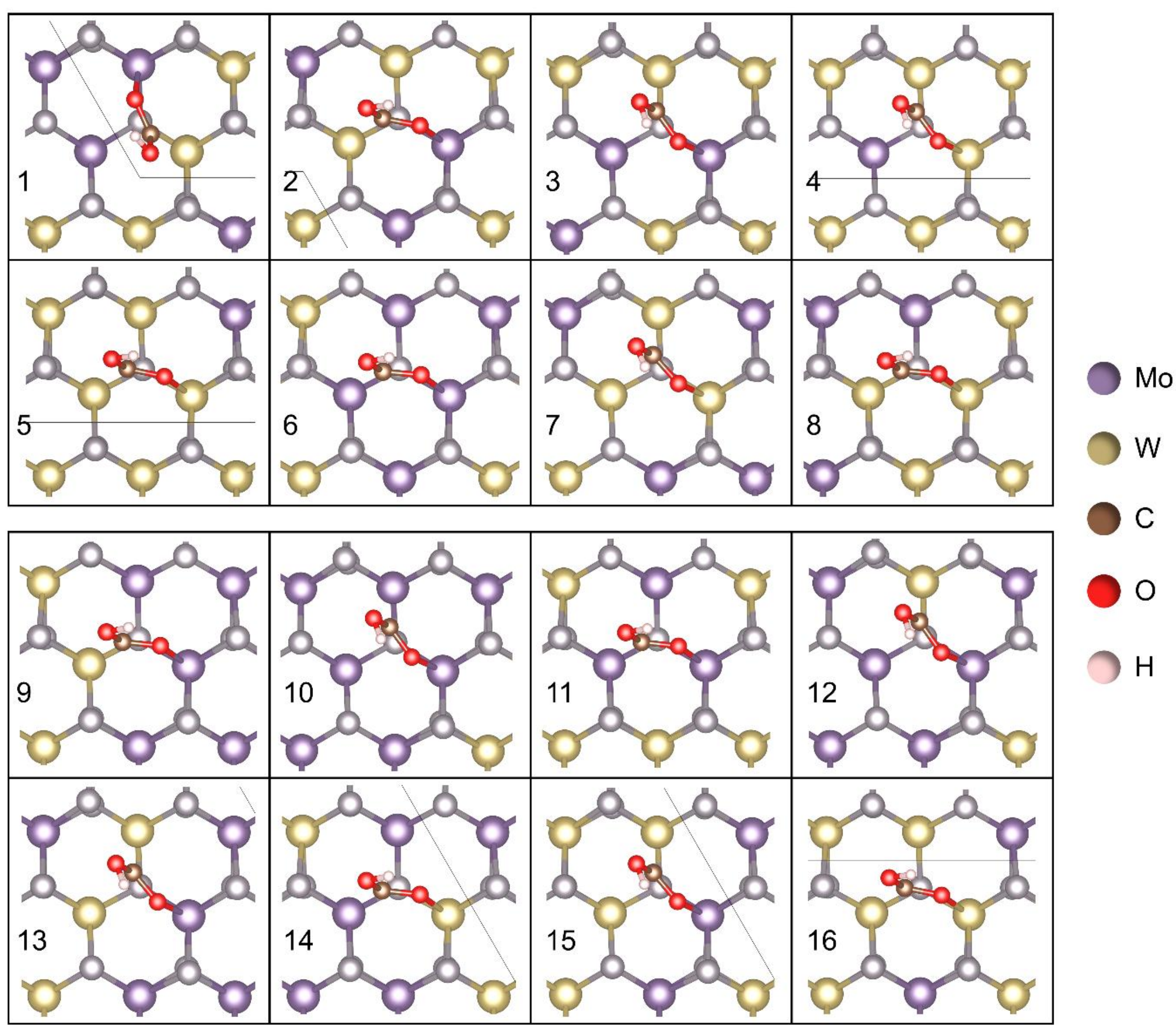

Figure S5. COOH* for every tested (Mo, W)$S_2$ local environment, each showing oxygen monodentate adsorption.

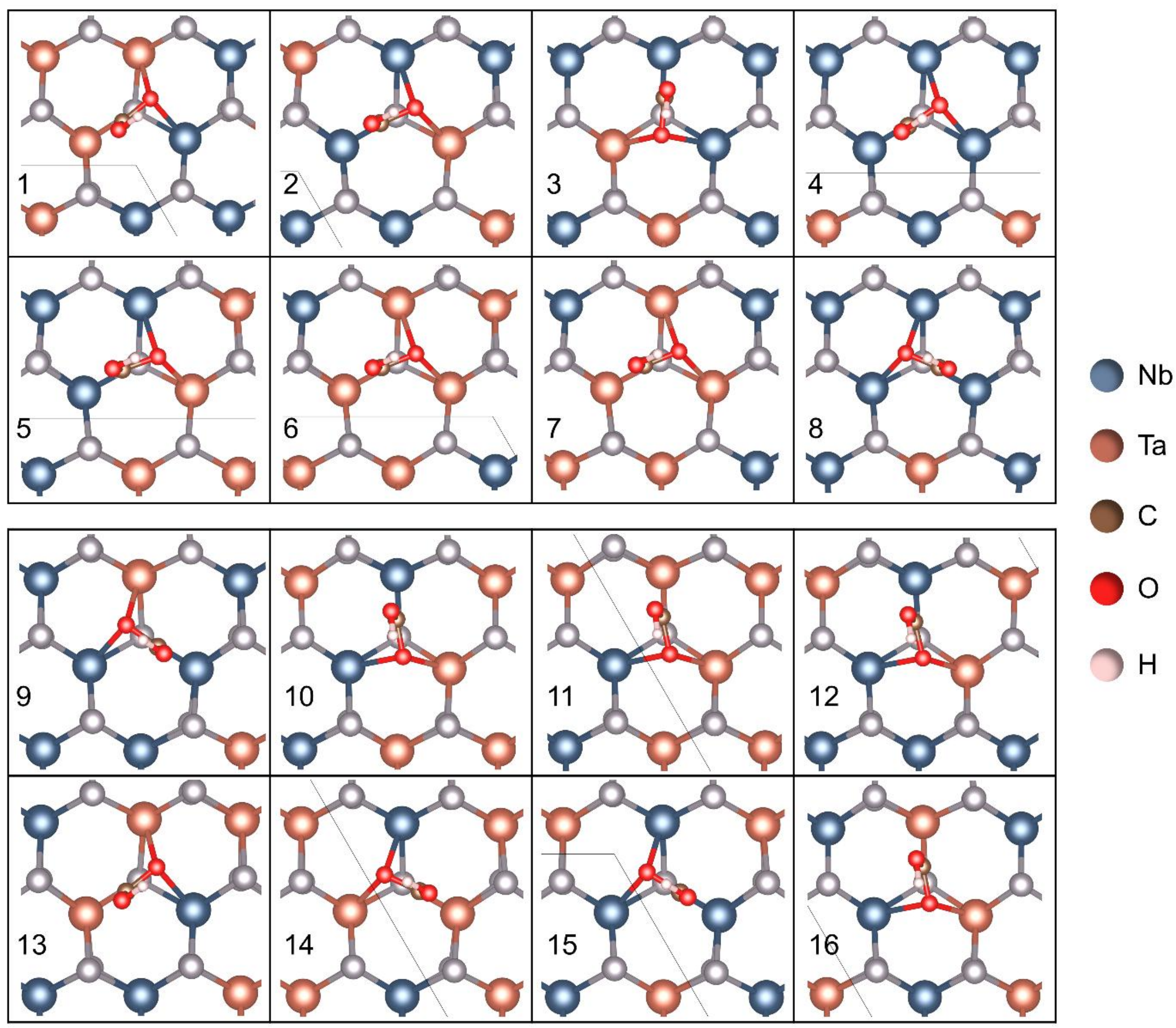

Figure S6. COOH* for every tested (Nb, Ta)$S_2$ local environment, each showing oxygen bidentate adsorption.

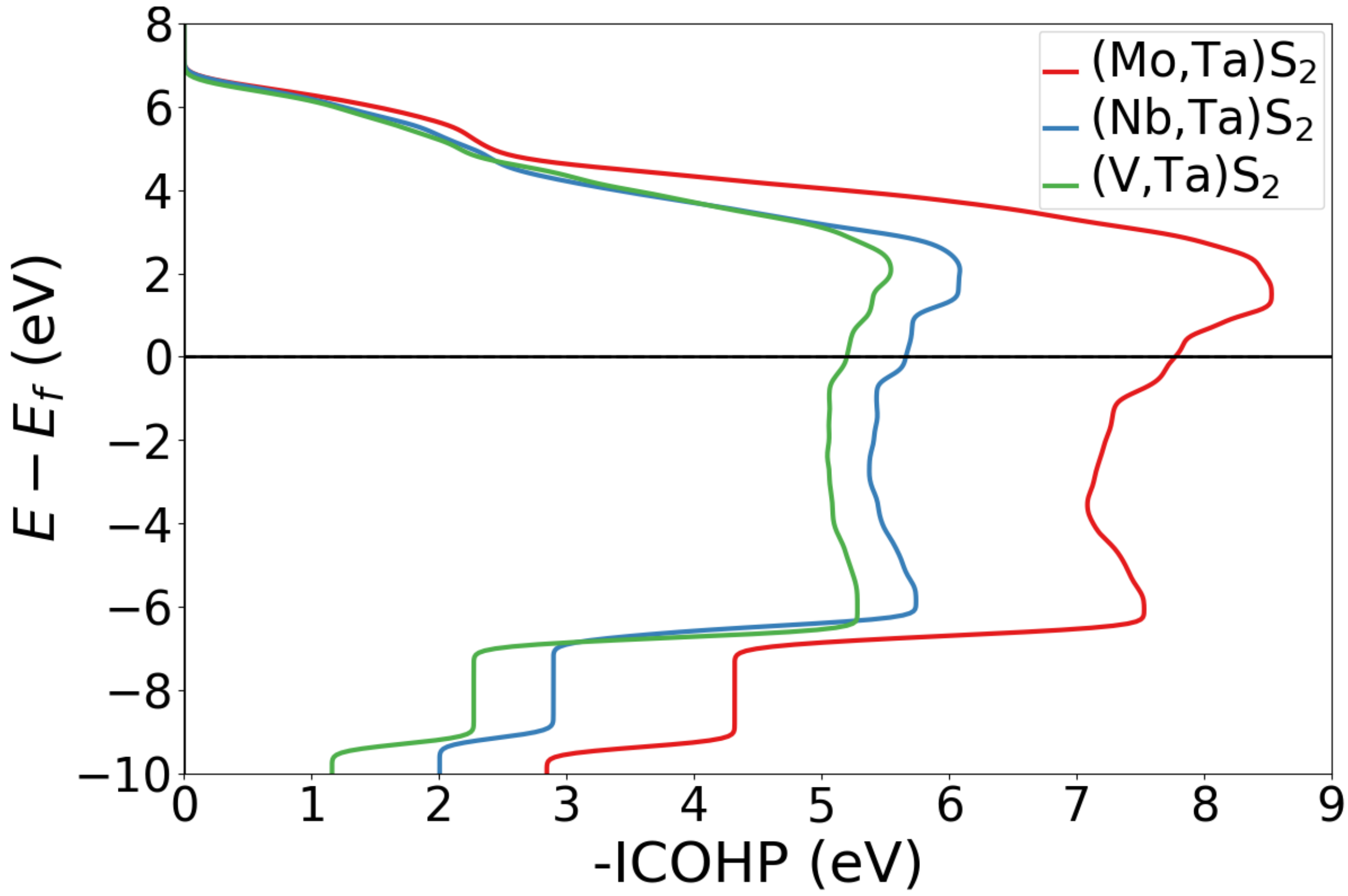

Figure S7. ICOHP of the three local environments analyzed in Fig. 5. The solid black line represents the Fermi energy.